\documentclass[12pt]{article}

\begin{document}

\begin{center}
{\Large\bf General relativistic tidal heating for the M$\o$ller 58
pseudotensor}\\
Lau Loi So ~ (26 Dec 2020)
\end{center}

\begin{abstract}
In his study of tidal stabilization of fully relativistic neutron
stars Thorne showed that the fully relativistic expression for
tidal heating is the same as in non-relativistic Newtonian theory.
Furthermore, Thorne also noted that this tidal heating must be
independent of how one localizes gravitational energy and is
unambiguously given by that expression. Favata calculated the
tidal heating for a number of classical gravitational
pseudotensors including that of M$\o$ller, and obtained the result
that all of them produced the same (Newtonian) value.  After a
re-examination of the calculation using the M$\o$ller pseudotensor
we find that indeed this pseudotensor gives the desired result
under the condition that the mass $M$ is a constant, while Favata
considered $M$ as being time dependent, which is illegitimate
since it violates the harmonic gauge condition, which he used.
Moreover, we carry on to consider this M$\o$ller pseudtensor even
in a black hole situation, i.e., beyond Newtonian physics.
\end{abstract}

\section{Introduction}
Dirac~\cite{Dirac} elucidated that it is not possible to obtain a
gravitational field energy expression that satisfies both
conditions: (1) when added to other forms of energy the total
energy is conserved, and (2) the energy within a definite
(three-dimensional) region at a certain time is independent of the
coordinate system.  For the classical pseudotensors, in general,
the first condition can be satisfied but the second not. In other
words, localizing gravitational energy is impossible.
Gravitational energy can be localized, however, there is no unique
proper coordinate independent way to localize it; instead we have
many expressions all reference frame dependent. Tidal heating is
an empirical physical phenomenon resulting from the net work done
by an external tidal field on an isolated body. The ocean tides on
Earth provide a familiar example of this kind of phenomenon.
However, a more dramatically example is the Jupiter-Io system,
where the moon Io's active volcanoes are the result of tidal
heating~\cite{Peale}.

In 1998 Thorne demonstrated that the expected tidal heating rate
is the same both in relativistic and Newtonian
gravity~\cite{Thorne,Purdue}:
$\dot{W}=-\frac{1}{2}\dot{I}_{ij}E^{ij}$, where $\dot{W}$ refers
to the work rate, the dot indicates the time derivative, $I_{ij}$
is the mass quadrupole moment of the isolated body and $E_{ij}$ is
the tidal field of the external universe.  Both $I_{ij}$ and
$E_{ij}$ are time dependent, symmetric and trace free. Moreover,
Thorne also noted that such tidal heating is independent of how
one localizes the gravitational energy and is unambiguously given
by a certain value. This has been verified by calculating
$\dot{W}$ explicitly using various gravitational pseudotensors to
represent the gravitational energy density and energy flux. In
contrast, there exists a recoverable process
$\dot{E}_{\rm{}int}\sim\frac{d}{dt}(I_{ij}E^{ij})$ which indicates
time reversible, where $E_{\rm{}int}$ is the energy interaction
between the isolated planet's quadrupolar deformation and the
external tidal field~\cite{Purdue}.

In 1999, Purdue used the Landau-Lifshitz pseudotensor to calculate
the tidal heating and confirmed that the result agreed with the
Newtonian perspective~\cite{Purdue,LL}. Later in 2001,
Favata~\cite{Favata} employed the same method to verify that the
Einstein, Bergmann-Thomson and M{\o}ller
pseudotensors~\cite{Freud,BT,Moller} give the same result as
Purdue found. Moreover, Booth and Creighton used the quasi-local
mass formalism of Brown and York to demonstrate the same
result~\cite{Booth}. All of them give the same value as the
Newtonian perspective. Referring to the work of Purdue and Favata,
this has completed the verification that the tidal heating is
indeed independent of the gravitational pseudotensor.

After a re-examination of the calculation for the M$\o$ller
pseudotensor, we noted that Favata misinterpreted the rate of
change of the constant mass $\dot{M}\neq0$. Here we claim that
$\dot{M}$ needs to be vanishing because of the harmonic gauge
requirement (see (\ref{3bMay2016}) and (\ref{21bDec2017}) below).
Nonetheless, Favata's and our result are the same eventually
(which we will define below). In the past, we thought that
obtaining the energy-momentum pseudotensor through a Freud type
superpotential (see (\ref{29aApril2016}) below) guarantees the
expected tidal heating~\cite{SoarXiv} but the converse is not
true. Surprisingly, the M$\o$ller pseudtensor provides a
counterexample. The present paper illustrates how the relativistic
tidal heating is indeed classical pseudotensor independent:
including the M$\o$ller pseudotensor. Explicitly, Thorne's
assertion is correct~\cite{Thorne}.

Our result shows that it does give the desired value for the
M$\o$ller pseudotensor. The nice property of the M$\o$ller
energy-momentum complex~\cite{Lessner} is that the energy content
of a hypersurface does not depend on the chosen spatial
coordinates, while the complexes proposed by Einstein,
Landau-Lifshitz, Bergmann-Thomson and Goldberg do. Perhaps this
may be the reason why there were many
investigators~\cite{Xulu,Yang,Radinschi,Nester1999} studying this
energy-momentum prescription in the past couple of decades. Thus
it is worthwhile to investigate the tidal heating using the
M$\o$ller pseudotensor.  After introducing the current quadrupole
moment $J^{ij}$ and magnetic type quadrupole moment
$B^{ij}$~\cite{Poisson,Zhang}, we step forward to consider the
black hole scenario using the M$\o$ller pseudotensor. Note that
$J_{ij}$ and $B_{ij}$ are time dependent, symmetric and traceless.

\section{Technical background}
We will use $\eta_{\mu\nu}=(-1,1,1,1)$ as our spacetime
signature~\cite{MTW} and the geometrical units $G=c=1$, where $G$
is the Newtonian gravitational constant and $c$ the speed of
light.  We adopt the convention that Greek letters indicate
spacetime indices and Latin letters refer to spatial indices. In
principle, the classical pseudotensors~\cite{CQGSoNesterChen2009}
can be obtained from a rearrangement of the Einstein equation:
$G_{\mu\nu}=\kappa{}T_{\mu\nu}$, where the constant
$\kappa=8\pi{}G/c^{4}$ and $T_{\mu\nu}$ is the material energy
tensor. This is a basic requirement for pseudotensors (see Ch. 20
in \cite{MTW}). One can define the gravitational energy-momentum
pseudotensor in terms of a suitable superpotential
$U_{\alpha}{}^{[\mu\nu]}$:
\begin{eqnarray}
2\kappa\sqrt{-g}t_{\alpha}{}^{\mu}:=\partial_{\nu}U_{\alpha}{}^{[\mu\nu]}
-2\sqrt{-g}G_{\alpha}{}^{\mu}.\label{6aMar2015}
\end{eqnarray}
The total energy-momentum density complex can then be defined as
\begin{eqnarray}
\sqrt{-g}{\cal{T}}_{\alpha}{}^{\mu}:=\sqrt{-g}(T_{\alpha}{}^{\mu}+t_{\alpha}{}^{\mu})
=(2\kappa)^{-1}\partial_{\nu}U_{\alpha}{}^{[\mu\nu]},
\end{eqnarray}
where to get the last equality we used (\ref{6aMar2015}) and the
Einstein equation. As a consequence of the antisymmetry of the
superpotential
$\partial_{\mu}(\sqrt{-g}{\cal{}T}_{\alpha}{}^{\mu})=0$. In vacuum
it leads to the energy conservation relations:
$\partial_{\mu}(\sqrt{-g}t_{0}{}^{\mu})=0$. The physical meaning
of $t_{0}{}^{0}$ and $t_{0}{}^{j}$ can be interpreted as the
gravitational energy density and energy flux. The energy-momentum
within a spatial region $V$ can be expressed as
$P_{\mu}=(-E,\vec{P})=\int_{V}\sqrt{-g}t_{\mu}{}^{0}d^{3}x$, the
sign for the Freud superpotential can be fixed by evaluating the
ADM mass~\cite{CQGSo2009,ADM}. Thus, tidal heating can be
manipulated as
\begin{eqnarray}
2\kappa\dot{W}=-\int_{V}\partial_{0}(\sqrt{-g}t_{0}{}^{0})d^{3}x
=\int_{V}\partial_{j}(\sqrt{-g}t_{0}{}^{j})d^{3}x.\label{21aSep2015}
\end{eqnarray}
Using Gauss's theorem, the last integral in (\ref{21aSep2015}) can
be converted into a surface integral of the form:
\begin{eqnarray}
2\kappa\dot{W}=\oint_{\partial{}V}\sqrt{-g}t_{0}{}^{j}dS_{j},\label{14aOct2015}
\end{eqnarray}
where $dS_{j}=\hat{n}_{j}r^{2}d\Omega$,
$\hat{n}_{j}\equiv{}x_{j}/r$ is the unit radial normal vector and
$r\equiv\sqrt{\delta_{ab}x^{a}x^{b}}$ is the distance from the
body in its local asymptotic rest frame.

For the tidal heating calculation, we adopt the harmonic gauge
\begin{eqnarray}
0=\partial_{\beta}(\sqrt{-g}g^{\alpha\beta})=-\sqrt{-g}\Gamma^{\alpha\beta}{}_{\beta}.
\end{eqnarray}
This harmonic coordinate condition provides the closest
approximation to rectilinear coordinates in curved space and is
suitable for studying gravitational waves~\cite{Dirac}. Decompose
the metric tensor out to 2nd order away from the Minkowski
background as follows~\cite{Zhang}:
\begin{eqnarray}
g_{\mu\nu}&=&\eta_{\mu\nu}+h_{\mu\nu}+h^{\alpha}{}_{\mu}h_{\alpha\nu}+k_{\mu\nu},\\
g^{\mu\nu}&=&\eta^{\mu\nu}-h^{\mu\nu}-k^{\mu\nu},
\end{eqnarray}
where
$k_{\mu\nu}=-\frac{1}{2}hh_{\mu\nu}+\frac{1}{8}\eta_{\mu\nu}h^{2}
-\frac{1}{4}\eta_{\mu\nu}h^{2}_{\alpha\beta}$. Here and after
indices are being transvected using the Minkowski
 background metric. However, it might not be so simple, as in the
 quantity $\Gamma^{\alpha\beta}{}_{\beta}$ the index $\beta$ is
 raised using the spacetime metric $g^{\alpha\beta}$. Note that we classify
$\eta_{\mu\nu}$ as the zeroth order, $h_{\mu\nu}$ the 1st order
and $k_{\mu\nu}$ the 2nd order. The trace of
$h:=\eta_{\alpha\beta}h^{\alpha\beta}$. The detailed expansion for
the harmonic gauge quantity is
\begin{eqnarray}
\Gamma^{\alpha\beta}{}_{\beta}=\partial_{\beta}\bar{h}^{\alpha\beta}+{\cal{O}}(h^{3}),\label{6aJuly2017}
\end{eqnarray}
where
$\bar{h}^{\alpha\beta}=h^{\alpha\beta}-\frac{1}{2}\eta^{\alpha\beta}h$.
In the following we will expand things out, keeping only the
relevant lowest order terms. We split the time and spatial
components of this gauge quantity as follows:
\begin{eqnarray}
0=\Gamma^{0\beta}{}_{\beta}=2\partial_{0}h^{00}+\partial_{c}h^{0c},\quad{}
0=\Gamma^{j\beta}{}_{\beta}=\partial_{0}h^{0j}+\partial_{c}\bar{h}^{jc}.\label{3bMay2016}
\end{eqnarray}
The related components (adopted from $(8)$ in~\cite{Zhang}) of the
gravitational field tensors are:
\begin{eqnarray}
h^{00}&=&\frac{2M}{r}+\frac{3}{r^{5}}I_{ab}x^{a}x^{b}-E_{ab}x^{a}x^{b},\label{10aSep2020}\\
h^{0j}&=&\frac{4}{r^{5}}\epsilon^{j}{}_{pq}J^{p}{}_{l}x^{q}x^{l}
+\frac{2}{3}\epsilon^{j}{}_{pq}B^{p}{}_{l}x^{q}x^{l}+\frac{2}{r^{3}}\dot{I}^{j}{}_{a}x^{a}
+\frac{2}{21}(5\dot{E}_{ab}x^{a}x^{b}x^{j}-2\dot{E}^{j}{}_{a}x^{a}r^{2}),\label{10bSep2020}\\
h^{ij}&=&\eta^{ij}h^{00}+\bar{h}^{ij},\label{19bJune2020}
\end{eqnarray}
where
$\bar{h}^{ij}=\frac{8}{3r^{3}}\epsilon_{pq}{}^{(i}\dot{J}^{j)p}x^{q}
+\frac{1}{21}\left[5x^{(i}\epsilon^{j)}{}_{pq}\dot{B}^{q}{}_{l}x^{p}x^{l}
-r^{2}\epsilon_{pq}{}^{(i}\dot{B}^{j)q}x^{p}\right]$ and
$\eta_{ij}\bar{h}^{ij}=0$. The value of the weighting factor
$\sqrt{-g}=1+h_{00}+{\cal{O}}(h^{2})$. From a calculation using
(\ref{10aSep2020}) and (\ref{10bSep2020}), we have
\begin{eqnarray}
2\partial_{0}h^{00}+\partial_{c}h^{0c}=4\dot{M}r^{-1}.\label{21bDec2017}
\end{eqnarray}
According to Thorne's argument the mass $M$ is constant in
time~\cite{Thorne} and indeed the harmonic gauge is valid at the
lowest order under the requirement that $\dot{M}$ has to be
vanishing. Explicitly, follows from (\ref{3bMay2016}):
$2\partial_{0}h^{00}+\partial_{c}h^{0c}=0$.

If the isolated body is absorbing an external quadrupolar field,
its mass quadrupole moment and current quadrupole moment
$\dot{I}_{ij}\neq0\neq\dot{J}_{ij}$ in general, and both of them
can generate tidal work. For our tidal heating calculation
purpose, we only pay attention to the lowest non-vanishing order;
to that order we will get a relation of the
form~\cite{Poisson,Zhang}
\begin{eqnarray}
\dot{W}=\partial_{0}(k_{1}I_{ij}E^{ij}+k_{2}J_{ij}B^{ij})
+k_{3}I_{ij}\dot{E}^{ij}+k_{4}J_{ij}\dot{B}^{ij},\label{23cSep2015}
\end{eqnarray}
where $k_{1},k_{2},k_{3},k_{4}$ are constants. The coefficients
$k_{1}$ and $k_{2}$ are related to a specific choice for the
energy localization, where $\partial_{0}(I_{ij}E^{ij})$ and
$\partial_{0}(J_{ij}B^{ij})$ are the ambiguous reversible
tidal-quadrupole interaction process. We expect to get
$(k_{3},k_{4})=(\frac{1}{2},\frac{2}{3})$ so that
$\frac{1}{2}I_{ij}\dot{E}^{ij}$ and
$\frac{2}{3}J_{ij}\dot{B}^{ij}$ are the unambiguous irreversible
tidal heating dissipation process that we are interested in.
Therefore we only look for the tidal heating coming from the
external tidal field $E_{ij}$ interacting with the evolving
quadrupole moment $I_{ij}$ of an isolated body. Similarly for
$J_{ij}$ and $B_{ij}$.

Here we consider the tidal heating due to the gravitational fields
$E_{ij}$ and $B_{ij}$ that come from the external universe.
Meanwhile there is a physical constraint that is assumed, the
Laplace equation $\vec{\nabla}^{2}(Mr^{-1})=0$. This means we
ignored a delta function at the origin that gives a non-vanishing
Poisson equation $\vec{\nabla}^{2}(Mr^{-1})\neq0$. Moreover, based
on the physical restriction of the harmonic gauge, we need the
mass $M$ to be time independent, no matter if it is in the
Jupiter-Io system or a black hole scenario. For the case of the
Jupiter-Io system, the tidal heating of Io is in a thermal
equilibrium state~\cite{Moore,Nature}, Io is absorbing
gravitational field energy from Jupiter and transfers the heat as
 convection. For the case of a black hole scenario, forms the extra
energy absorbed from the outside universe will transfer to other
forms of energy such as gravitational waves, thus the mass $M$
remains constant.

\section{Tidal heating from the Freud superpotential}
There are an infinite number of superpotentials, the
Freud~\cite{Freud} superpotential $_{F}U_{\alpha}{}^{[\mu\nu]}
:=-\sqrt{-g}g^{\beta\sigma}\Gamma^{\tau}{}_{\beta\lambda}
\delta^{\lambda\mu\nu}_{\tau\sigma\alpha}$ is a straightforward
expression that can be used for illustrating the tidal work.  Here
we use it to reproduce the result of the tidal heating for the
Einstein pseudotensor. We have mentioned that the energy-momentum
complex can be computed as
$\sqrt{-g}({}_{E}{\cal{T}}_{\alpha}{}^{\mu})=(2\kappa)^{-1}\partial_{\nu}({}_{F}U_{\alpha}{}^{[\mu\nu]})$.
As we will explain in more detail a little further below, at any
point, to lowest order in Riemann normal coordinates inside matter
this reduces to the source energy-momentum stress tensor
$T_{\alpha}{}^{\mu}=\kappa^{-1}G_{\alpha}{}^{\mu}$, as one expects
from the equivalence principle. In vacuum, the Einstein
pseudotensor is~\cite{CQGSoNesterChen2009}
\begin{eqnarray}
2\kappa({}_{E}t_{\alpha}{}^{\mu})
&=&\delta^{\mu}_{\alpha}(\Gamma^{\beta\lambda}{}_{\nu}\Gamma^{\nu}{}_{\beta\lambda}
-\Gamma^{\pi}{}_{\pi\nu}\Gamma^{\nu\lambda}{}_{\lambda})
+\Gamma^{\nu}{}_{\beta\nu}(\Gamma^{\mu\beta}{}_{\alpha}+\Gamma^{\beta\mu}{}_{\alpha})
+\Gamma^{\pi}{}_{\pi\alpha}(\Gamma^{\mu\lambda}{}_{\lambda}-\Gamma^{\lambda\mu}{}_{\lambda})\nonumber\\
&&-2\Gamma^{\beta\nu}{}_{\alpha}\Gamma^{\mu}{}_{\beta\nu}.
\end{eqnarray}
In general, this pseudotenosr is not symmetric. In the harmonic
gauge, the gravitational energy density and energy
flux~\cite{Favata} are
\begin{eqnarray}
2\kappa({}_{E}t_{0}{}^{0})=-\frac{1}{2}\eta^{cd}h_{00,c}h_{00,d},\quad{}
2\kappa({}_{E}t_{0}{}^{j})=\eta^{ja}h_{00,0}h_{00,a}.\label{22aOct2015}
\end{eqnarray}
Using (\ref{22aOct2015}b) in (\ref{14aOct2015}), we recover the
known result
$\dot{W}_{E}=\frac{3}{10}\partial_{0}(I_{ij}E^{ij})-\frac{1}{2}\dot{I}_{ij}E^{ij}$
as Favata obtained.

To the  order of concern here, it is sufficient to consider
superpotentials that are linear in the connection. There are only
three possible terms with suitable symmetry, one by itself is
proportional to the M$\o$ller superpotential. In principle, the
three parameter expression can be cast in the following way:
\begin{eqnarray}
U_{\alpha}{}^{[\mu\nu]}:=\sqrt{-g}(
a_{1}\delta^{\tau}_{\alpha}\Gamma^{\rho\lambda}{}_{\lambda}
+a_{2}\delta^{\rho}_{\alpha}\Gamma^{\lambda\tau}{}_{\lambda}
+a_{3}\Gamma^{\tau\rho}{}_{\alpha})\delta^{\mu\nu}_{\rho\tau},\label{29aApril2016}
\end{eqnarray}
where $a_{1}, a_{2}, a_{3}$ are real. One limit that should be
considered is the small region limit. Around any arbitrary point,
one can introduce Riemann normal
coordinates~\cite{CQGSo2009,SoNesterPRD} with the origin at that
point, such that
\begin{equation}
g_{\alpha\beta}|_{0}=\eta_{\alpha\beta},\quad{}
g_{\alpha\beta,\mu}|_{0}=0,\quad{}
-3\Gamma^{\alpha}{}_{\beta\mu,\nu}|_{0}=R^{\alpha}{}_{\beta\mu\nu}+R^{\alpha}{}_{\mu\beta\nu}.
\end{equation}
According to the equivalence principle, to lowest order the
pseudotensor associated with the above superpotential should
reduce to the source energy-momentum; for the superpotential
(\ref{29aApril2016}) this requirement yields
\begin{eqnarray}
2\kappa{}T_{\alpha}{}^{\mu}=\frac{1}{3}\left[
(2a_{1}+a_{2}+3a_{3})R_{\alpha}{}^{\mu}-(2a_{1}+a_{2})\delta^{\mu}_{\alpha}R\right].
\end{eqnarray}
In order for this to agree with the Einstein field equation, we
have the following constraints~\cite{CQGSo2009}:
\begin{eqnarray}
2a_{1}+a_{2}=3,\quad{}a_{3}=1.\label{27aSep2020}
\end{eqnarray}
Now consider the linearized field equation, up to the first order
$g_{\alpha\beta}\simeq\eta_{\alpha\beta}+h_{\alpha\beta}$, we have
the following
\begin{eqnarray}
2\kappa{}T_{\alpha}{}^{\mu}&=&a_{1}[\partial^{2}_{\alpha\beta}\bar{h}^{\beta\mu}
-\delta^{\mu}_{\alpha}(\partial^{2}_{\beta\lambda}\bar{h}^{\beta\lambda})]
+a_{3}\partial^{\mu}\partial_{\beta}\bar{h}_{\alpha}{}^{\beta}\nonumber\\
&&+\frac{1}{2}(a_{3}-a_{2})\partial_{\alpha}\partial^{\mu}h
-\partial^{\lambda}\partial_{\lambda}\left(a_{3}h_{\alpha}{}^{\mu}
-\frac{1}{2}a_{2}\delta^{\mu}_{\alpha}h\right).\label{29aSep2020}
\end{eqnarray}
The standard approach requires that the field equation reduces to
the wave equations
$\frac{1}{2}\partial^{\lambda}\partial_{\lambda}\bar{h}_{\alpha}{}^{\mu}=-\kappa{}T_{\alpha}{}^{\mu}$
under the criterion that the harmonic gauge is chosen. Referring
to (\ref{29aSep2020}), the constraints are
\begin{eqnarray}
a_{1}={\mbox{arbitrary}},\quad{}a_{2}=a_{3}.\label{27bSep2020}
\end{eqnarray}
Combining (\ref{27aSep2020}) and (\ref{27bSep2020}), we have the
unique solution $a_{1},a_{2},a_{3}$ are all unity.

Here we explain what we mean by a Freud type superpotential:
decompose the Freud superpotential as follows
\begin{eqnarray}
_{F}U_{\alpha}{}^{[\mu\nu]}:=-\sqrt{-g}(\eta^{\beta\sigma}-h^{\beta\sigma})
\Gamma^{\tau}{}_{\beta\lambda}\delta^{\lambda\mu\nu}_{\tau\sigma\alpha}.
\label{15aOct2015}
\end{eqnarray}
Note that the linear in $\eta\Gamma$ terms give the expected
interior mass and tidal heating, while the $h\Gamma$ terms only
alter the value $\partial_{0}(I_{ij}E^{ij})$ or
$\partial_{0}(J_{ij}B^{ij})$~\cite{SoarXiv}. Any superpotential
that agrees with the Freud superpotential to lowest order in
$h_{\mu\nu}:=g_{\mu\nu}-\eta_{\mu\nu}$, is referred to as a Freud
type superpotential. We know the Landau-Lifshitz (LL)
superpotential can be identified as a Freud type superpotential
since we can raise the indices:
$_{LL}U^{\alpha[\mu\nu]}={}_{F}U_{\beta}{}^{[\mu\nu]}\sqrt{-g}g^{\alpha\beta}$,
which gives the desired interior mass and tidal work, whereas the
extra weighting factor $\sqrt{-g}$, once again, only affects
$\partial_{0}(I_{ij}E^{ij})$ or $\partial_{0}(J_{ij}B^{ij})$.
Another example is the Papapetrou superpotential~\cite{SoarXiv}:
\begin{eqnarray}
_{P}U^{\alpha[\mu\nu]}={}_{F}U_{\beta}{}^{[\mu\nu]}g^{\alpha\beta}
-\sqrt{-g}(g^{\rho\tau}h^{\pi\gamma}\Gamma^{\sigma}{}_{\lambda\pi}
+g^{\rho\pi}h^{\sigma\tau}\Gamma^{\gamma}{}_{\lambda\pi})
\delta_{\tau\gamma}^{\mu\nu}\delta_{\rho\sigma}^{\lambda\alpha}.
\end{eqnarray}
Referring to (\ref{29aApril2016}), to lowest order there are just
three possible superpotential terms and each term has its own
characteristic features.

\subsection{The 1st term of the Freud superpotential}
When $(a_{1},a_{2},a_{3})=(1,0,0)$ for (\ref{29aApril2016}), the
first term of the Freud type superpotential is
$_{1}U_{\alpha}{}^{[\mu\nu]}:=\sqrt{-g}\Gamma^{\rho\lambda}{}_{\lambda}\delta^{\mu\nu}_{\rho\alpha}$.
The corresponding energy-momentum complex is
\begin{eqnarray}
(2\kappa){}_{1}{\cal{T}}_{\alpha}{}^{\mu}
=(\partial_{\nu}+\Gamma^{\pi}{}_{\pi\nu})\Gamma^{\rho\lambda}{}_{\lambda}\delta^{\mu\nu}_{\rho\alpha}.
\label{15aApril2016}
\end{eqnarray}
Inside matter at the origin, the energy-momentum
$_{1}{\cal{T}}_{\alpha}{}^{\mu}=(R_{\alpha}{}^{\mu}-\delta^{\mu}_{\alpha}R)/(3\kappa)$
in Riemann normal coordinates.  Upon applying the harmonic gauge
in vacuum, the pseudotensor $_{1}t_{\alpha}{}^{\mu}=0$, i.e., both
the energy density $_{1}t_{0}{}^{0}$ and energy flux
$_{1}t_{0}{}^{j}$ are zero.

\subsection{The 2nd term of the Freud superpotential}
When $(a_{1},a_{2},a_{3})=(0,1,0)$ for (\ref{29aApril2016}), the
second term of the Freud type superpotential is
$_{2}U_{\alpha}{}^{\mu\nu}:=\sqrt{-g}\delta^{\mu\nu}_{\alpha\tau}\Gamma^{\lambda\tau}{}_{\lambda}$.
The associated energy-momentum complex is
\begin{eqnarray}
(2\kappa){}_{2}{\cal{T}}_{\alpha}{}^{\mu}
=\delta^{\mu}_{\alpha}(-R+\partial_{\lambda}\Gamma^{\lambda\beta}{}_{\beta}
+\Gamma^{\beta\nu}{}_{\lambda}\Gamma^{\lambda}{}_{\beta\nu})
-(\partial_{\alpha}+\Gamma^{\pi}{}_{\pi\alpha})\Gamma^{\lambda\mu}{}_{\lambda}.
\end{eqnarray}
Inside matter,
$_{2}{\cal{T}}_{\alpha}{}^{\mu}=(R_{\alpha}{}^{\mu}-\delta^{\mu}_{\alpha}R)/(6\kappa)$
to lowest order in Riemann normal coordinates. In vacuum, using
the harmonic gauge condition, this $S$ pseudotensor can be written
as
\begin{eqnarray}
(2\kappa){}_{2}t_{\alpha}{}^{\mu}
=\delta^{\mu}_{\alpha}\Gamma^{\beta\nu}{}_{\lambda}\Gamma^{\lambda}{}_{\beta\nu}
-(\partial_{\alpha}+\Gamma^{\pi}{}_{\pi\alpha})\Gamma^{\lambda\mu}{}_{\lambda}.
\end{eqnarray}
The related gravitational energy density is
\begin{eqnarray}
(2\kappa){}_{2}t_{0}{}^{0}
&=&(1-h_{00})\partial^{2}_{00}h_{00}-\frac{3}{2}(\partial_{0}h_{00})^{2}
+h^{0c}\partial^{2}_{0c}h_{00}-\partial_{0}(h^{0c}\partial_{0}h_{0c})\nonumber\\
&&-(\partial_{c}h_{00})(\partial_{0}\bar{h}^{0c})
-(\partial_{0}\bar{h}^{cd})(\partial_{c}h_{0d})
-\frac{1}{2}\bar{h}^{cd}(\partial^{2}_{00}\bar{h}_{cd})\nonumber\\
&&+\frac{1}{2}\left[(\partial^{c}h^{0d})(\partial_{d}h_{0c})
-(\partial^{c}h_{00})(\partial_{c}h_{00})
-(\partial^{c}h^{0d})(\partial_{c}h_{0d})\right]\nonumber\\
&&+\frac{1}{2}(\partial^{c}\bar{h}^{de})(\partial_{d}\bar{h}_{ce})
-\frac{1}{4}\left[(\partial_{0}\bar{h}^{cd})(\partial_{0}\bar{h}_{cd})
+(\partial^{c}\bar{h}^{de})(\partial_{c}\bar{h}_{de})\right],
\end{eqnarray}
and the energy flux is
\begin{eqnarray}
(2\kappa)\,_{2}t_{0}{}^{j}&=&-\partial_{0}\partial^{j}h_{00}+(\partial_{0}h_{00})\,(\partial_{0}h^{0j})
+(\partial_{0}\bar{h}^{ja})(\partial_{a}h_{00})
+h^{0j}\partial^{2}_{00}h_{00}\nonumber\\
&&+\bar{h}^{ja}\partial^{2}_{0a}h_{00}
+\partial_{0}\left[2h_{00}\partial^{j}h_{00}+h^{0c}\partial^{j}h_{0c}
+\frac{1}{2}\bar{h}^{cd}\partial^{j}\bar{h}_{cd}\right].
\end{eqnarray}
According to (\ref{21aSep2015}) and (\ref{14aOct2015}), compute
the tidal heating as follows
\begin{eqnarray}
\dot{W}_{2}&=&\frac{1}{2\kappa}\frac{d}{dt}\left[
-\int_{V}\vec{\nabla}^{2}h_{00}d^{3}x +\oint_{\partial{}V}\left(
2h_{00}\partial^{j}h_{00}+h^{0c}\partial^{j}h_{0c}\right)dS_{j}\right]\nonumber\\
&=&\frac{1}{5}\frac{d}{dt}\left(I_{ij}E^{ij}+\frac{2}{3}J_{ij}B^{ij}\right),
\end{eqnarray}
where $\vec{\nabla}^{2}h_{00}=2\vec{\nabla}^{2}(Mr^{-1})=0$ as
mentioned before. This part contributes vanishing tidal heating
since it is undertaken a time reversible process. In particular,
Purdue uses $W_{\rm{}int}=\frac{\gamma+2}{10}I_{ij}E^{ij}$ to
interpret different choices of energy localization by tuning the
coefficient $\gamma$~\cite{Purdue}.

\subsection{The 3rd term of the Freud superpotential}
When $(a_{1},a_{2},a_{3})=(0,0,1)$ for (\ref{29aApril2016}), the
third term of the Freud type superpotential is
$_{3}U_{\alpha}{}^{[\mu\nu]}:=\sqrt{-g}\Gamma^{\tau\rho}{}_{\alpha}\delta^{\mu\nu}_{\rho\tau}$.
If we multiply by a factor of 2, the M$\o$ller
superpotential~\cite{Moller} is recovered. This is the essential
part which gives the desired tidal heating value. The associated
M$\o$ller pseudotensor can be obtained as
\begin{eqnarray}
2\kappa({}_{3}t_{\alpha}{}^{\mu})
=2R_{\alpha}{}^{\mu}-\partial_{\alpha}\Gamma^{\mu\beta}{}_{\beta}
+g^{\beta\mu}\partial_{\alpha}\Gamma^{\nu}{}_{\beta\nu}
-2\Gamma^{\beta\nu}{}_{\alpha}\Gamma^{\mu}{}_{\beta\nu}.\label{12aSep2020}
\end{eqnarray}
Inside matter this reduces to the total energy-momentum to lowest
order is
$_{3}{\cal{T}}_{\alpha}{}^{\mu}=R_{\alpha}{}^{\mu}/(2\kappa)$ at
the origin in Riemann normal coordinates. Since this result is not
compatible with Einstein's equation, one may have doubts as to
whether it is meaningful to keep calculating the tidal work?
Although the M$\o$ller pseudotensor has already failed the inside
matter requirement, this pseudotensor has the nice feature that
its gravitational energy is coordinate system independent. In
vacuum, referring to (\ref{12aSep2020}), the M$\o$ller
pseudotensor becomes
\begin{eqnarray}
2\kappa(_{3}t_{\alpha}{}^{\mu})=-\partial_{\alpha}\Gamma^{\mu\beta}{}_{\beta}
+\frac{1}{2}(\partial_{\alpha}\partial^{\mu}h-h^{\beta\mu}\partial^{2}_{\alpha\beta}h
-h^{\beta\lambda}\partial_{\alpha}\partial^{\mu}h_{\beta\lambda})
-(\partial_{\alpha}h^{\beta\lambda})(\partial_{\beta}h^{\mu}{}_{\lambda}).\label{12bSep2020}
\end{eqnarray}
The associated energy density and energy flux are
\begin{eqnarray}
2\kappa({}_{3}t_{0}{}^{0})&=&(h_{00}-1)\partial^{2}_{00}h_{00}+3(\partial_{0}h_{00})^{2}
+(\partial_{c}h_{0d})(\partial_{0}\bar{h}^{cd})+\frac{1}{2}\bar{h}^{cd}\partial^{2}_{00}\bar{h}_{cd}
\nonumber\\
&&+\eta^{cd}\left[h_{0c}(\partial^{2}_{0d}h_{00}
-\partial^{2}_{00}h_{0d})-(\partial_{0}h_{0c})(\partial_{0}h_{0d}+\partial_{d}h_{00})\right],\\
2\kappa({}_{3}t_{0}{}^{j})&=&(1-3h_{00})\partial_{0}\partial^{j}h_{00}
-(\partial_{0}h_{00})(\partial^{j}h_{00})-h^{0j}\partial^{2}_{00}h_{00}\nonumber\\
&&-\partial_{c}(\bar{h}^{jc}\partial_{0}h_{00})
+(\partial_{0}h_{0c}-\partial_{c}h_{00})(\partial_{0}\bar{h}^{jc})
-(\partial_{0}\bar{h}_{cd})(\partial^{c}\bar{h}^{jd})\nonumber\\
&&+\eta^{cd}\left[h_{0c}\partial_{0}\partial^{j}h_{0d}-(\partial_{0}h_{0c})(\partial_{d}h^{0j})\right]
-\frac{1}{2}\bar{h}_{cd}\partial_{0}\partial^{j}\bar{h}^{cd}.
\end{eqnarray}
Calculate the tidal heating as follows
\begin{eqnarray}
2\kappa\dot{W}_{3}&=&\frac{d}{dt}\int_{V}\vec{\nabla}^{2}h_{00}d^{3}x
-\oint_{\partial{}V}\left[2h_{00}\partial_{0}\partial^{j}h_{00}
+(\partial_{0}h_{00})(\partial^{j}h_{00})\right]dS_{j}\nonumber\\
&&+\oint_{\partial{}V}\eta^{cd}\left[h_{0c}\partial_{0}\partial^{j}h_{0d}
-(\partial_{0}h_{0c})(\partial_{d}h^{0j})\right]dS_{j}.
\end{eqnarray}
Again, we emphasize that there is a difference between Favata's
and our understanding. The mass $M$ is time dependent in Favata's
argument, but we treated $M$ as a constant, based on the harmonic
gauge constraint (see (\ref{3bMay2016}) and (\ref{21bDec2017})).

After some simple algebra, the tidal heating for a black hole is
\begin{eqnarray}
\dot{W}_{3}&=&\frac{1}{2}\left(I_{ij}\dot{E}^{ij}+\frac{4}{3}\,J_{ij}\dot{B}^{ij}\right)
-\frac{2}{5}\frac{d}{dt}\left(I_{ij}E^{ij}+\frac{4}{3}J_{ij}B^{ij}\right)\nonumber\\
&=&\frac{16}{45}M^{6}(\dot{E}^{2}_{ij}+\dot{B}^{2}_{ij})
-\frac{2}{5}\frac{d}{dt}\left(I_{ij}E^{ij}+\frac{4}{3}J_{ij}B^{ij}\right),
\end{eqnarray}
where we used~\cite{Poisson}:
\begin{eqnarray}
I_{ij}=\frac{32M^{6}}{45}\,\dot{E}_{ij},\quad{}J_{ij}=\frac{8M^{6}}{15}\dot{B}_{ij}.
\end{eqnarray}
Note that $E^{2}_{ij}$ means $E^{ij}E_{ij}$ and similarly for
$B^{2}_{ij}$. Only this superpotential is the substantial  part
which contributes the desired tidal heating.  More accurately,
besides failing to meet the inside matter result
$2G^{\mu}{}_{\nu}$, we discovered that whenever the superpotential
includes this term with unit magnitude, one can guarantee that the
suitable tidal heating value can be achieved. In other words, the
tidal heating is (in a suitable sense) pseudotensor independent as
Thorne expected and Favata intended to verify~\cite{Favata}.
Thorne wrote: ``Similarly, if, in our general relativistic
analysis, we were to change our energy localization by switching
from the Landau-Lifshitz pseudotensor to some other pseudotensor,
or by performing a gauge change on the gravitational field, we
thereby would alter $E_{\rm{}int}$ but leave $W$ unchanged"~(p.9
in \cite{Thorne}). Perhaps Thorne had assumed that all
pseudotensors already had the standard form to linear order (see
Ch. 20 in \cite{MTW}). From our result it turns out that the
M$\o$ller pseudotensor (which \emph{does not} fit into the
standard linear form) also fails to yield the standard tidal
heating. In other words, all the Freud type classical
pesudotensors yield the expected tidal heating value, which shows
that Thorne's assertion is correct~\cite{Thorne}.

The M$\o$ler superpotential is twice the 3rd Freud superpotential term, 
giving the consequent small region M$\o$ller result and the Io tidal heating.  
There two discoveries here: the 3rd Freud superpotentail term gives the desired value, 
and the M$\o$ller superpotential does not.

\section{Conclusion}
Thorne argued that tidal heating is independent of how one
localizes the gravitational energy and the value is unambiguous.
Purdue and Favata used a number of well known pseudotensors to
calculate the tidal heating and verified that Thorne's assertion
is correct for them. After a re-examination of the M$\o$ller
pseudotensor, we found reasons to doubt Favata's calculations.
Substantially, a technical difference had arisen. For the mass we
used $\dot{M}=0$ which means that $M$ is a constant, while Favata
treated $\dot{M}\neq0$ which requires $M$ is a time dependent
object. However, it is strictly forbidden to allow $M$ as a time
dependent function according to the harmonic gauge. Moreover, our
result is valid both in the Jupiter-Io type system and black hole
scenario.

Here we emphasize that if a suitable gravitational energy-momentum
pseudotensor fulfills the Freud type superpoential condition, this
requirement ensures the expected tidal heating. Our analysis
indicates that the Freud type superpotential is sufficient but not
necessary. In particular, the pseudotensor that is obtained from
1/2 the M$\o$ller superpotential gives a counterexample that
succeeds in achieving the desired tidal heating, even though it
has the physical handicap of failing to meet the inside matter
requirement.

\section*{Acknowledgment}
The author would like to thank Dr. Peter Dobson, Professor
Emeritus, HKUST, for reading the manuscript and providing some
helpful comments.

\end{document}